\documentclass[12pt]{article}
\usepackage{authblk}

\usepackage[utf8]{inputenc}
\usepackage{environ}  

\NewEnviron{keyword}{%
  \par
}
\usepackage{amsmath, amssymb, amsfonts, amsthm, latexsym, mathrsfs}

\usepackage{tikz}
\usepackage{tikz-cd}
\usetikzlibrary{arrows.meta, decorations.pathreplacing, arrows}
\usepackage{pgfplots}

\usepackage{array}
\usepackage{longtable}
\usepackage{booktabs}
\usepackage{tabularx}

\usepackage{listings}
\usepackage{algorithmic}

\usepackage{caption}
\usepackage{subcaption}
\usepackage{float}

\usepackage{import}
\usepackage{nopageno}
\usepackage{hyperref}
\usepackage{url}
\usepackage{doi}

\usepackage{sidecap}
\usepackage{todonotes}

\usepackage[margin=1in]{geometry}

\usepackage[numbers]{natbib}

\theoremstyle{plain}

\theoremstyle{definition}

\theoremstyle{plain}



\title{Mathematical Insights into Protein Architecture: Persistent Homology and Machine Learning Applied to the Flagellar Motor}

\author[1]{Zakaria~Lamine\thanks{Corresponding author: \texttt{z.lamine@uit.ac.ma}}}
\author[2]{Abdelatif~Hafid}
\author[3]{Mohamed~Rahouti}

\affil[1]{Department of Mathematics, Faculty of Sciences, Ibn Tofail University, Kenitra, Morocco}
\affil[2]{Department of Mathematics, Faculty of Sciences and Techniques, Abdelmalek Essaâdi University, Al Hoceima, Morocco}
\affil[3]{Department of Computer and Information Science, Fordham University, New York, NY 10023, USA}

\begin{document}

\maketitle

\begin{abstract}
We present a machine learning approach that leverages persistent homology to classify bacterial flagellar motors into two functional states: rotated and stalled. By embedding protein structural data into a topological framework, we extract multiscale features from filtered simplicial complexes constructed over atomic coordinates. These topological invariants, specifically persistence diagrams and barcodes, capture critical geometric and connectivity patterns that correlate with motor function. The extracted features are vectorized and integrated into a machine learning pipeline that includes dimensionality reduction and supervised classification. Applied to a curated dataset of experimentally characterized flagellar motors from diverse bacterial species, our model demonstrates high classification accuracy and robustness to structural variation. This approach highlights the power of topological data analysis in revealing functionally relevant patterns beyond the reach of traditional geometric descriptors, offering a novel computational tool for protein function prediction.
\end{abstract}



\section{Introduction}
\label{sec:introduction}
Proteins are fundamental biological macromolecules that perform various functions in living organisms, including enzymatic catalysis, signal transduction, and structural support \citep{b25}. The three-dimensional shape of a protein determines its functionality, making the study of protein structure a cornerstone of molecular biology \citep{b26}.

Recent advances in high-throughput technologies, such as X-ray crystallography, cryo-electron microscopy, and deep sequencing, have generated vast amounts of structural and functional data. However, analyzing and interpreting this data remains a significant challenge due to the complexity of protein structures and their dynamic nature \citep{b22}\citep{b29}\citep{b30}.

The use of topological methods, particularly persistent homology, has gained traction in the study of protein structures \citep{b24}. Persistent homology has been applied to identify critical features of protein folding; Analyze binding sites and active regions of enzymes and understand dynamic conformational changes in proteins. 

Moreover, machine learning approaches have been integrated with topological descriptors, demonstrating significant potential in predicting protein functionality and stability \citep{b21}\citep{b23}. Despite these advances, the application of more refined tools, such as persistent homology computations, remains underexplored \citep{b0}. Although persistent homology effectively captures global and locomotive characteristics, it focuses primarily on \emph{presence} of topological characteristics such as connected components, loops, and voids \citep{b24}. However, it often neglects additional algebraic structures, such as persistent homology classes, which provide richer invariants to understand interactions within protein structures \citep{b28}.

Furthermore, there is limited exploration of how persistent homology computations can improve protein structure classification and provide new information on the relationships between structural and functional properties \citep{b27}, as well as seamless integration with statistical and machine learning pipelines for predictive analysis.

This study aims to improve protein analysis by introducing persistent homology methods, addressing a key limitation in current approaches. Specifically, we used free resolutions and computed operators to derive persistent homology data that provide deeper insights into protein structures. We demonstrate the mathematical model by focusing on the Flagellar Motor using this biologically significant protein complex as a case study. In doing so, we explore how persistent homology computations can reveal structural and functional relationships in molecular machinery. Finally, this paper's key contribution is to advance Topological Data Analysis (TDA) by introducing a novel framework for categorifying topological invariants in the context of biological data. This is accomplished through an algebraic topological perspective.

The paper is structured as follows. Section \ref{sec:mathematical_model} establishes the theoretical framework for persistent homology computation and presents a rigorous formulation of the model's algebraic topological characteristics. Section \ref{sec:Methods} presents the implementation methodology using a well-established protein folding dataset, immediately section \ref{sec:Results} follows to summarize the obtained results. Section \ref{sec:Discussions} is devoted to a comparative analysis. Finally,  Section \ref{sec:conclusion} concludes the paper.

\section{Mathematical Model}
\label{sec:mathematical_model}

Traditional methods in protein structure analysis, such as Root Mean Square Deviation (RMSD), sequence alignment, and energy-based modeling, have been effective for tasks such as structure comparison, functional annotation, and fold prediction. However, these approaches often fall short in capturing the intricate, higher-dimensional relationships inherent in protein structures. For instance; They are scalar measures that depend on global alignment or pointwise distances, which can overlook essential structural motifs, such as cavities or loops. They focus on local residue matches, often missing broader topological features related to protein function. In the other side; Energy-based models excel at evaluating pairwise interactions but struggle to generalize to multi-dimensional features, such as collective binding sites or channels. persistent homology, in contrast, provides a higher-dimensional view of these structures, characterizing features like voids (via \( H^2 \)) and connectivity patterns (via \( H^1 \)); it captures invariants that persist across structural deformations, providing a robust framework to identify conserved functional regions independent of local perturbations; inherently encodes such features, offering insights into protein-ligand interactions and active site geometry; its invariants are naturally robust to noise and minor perturbations, making them ideal for analyzing experimental structures with uncertainties. Moreover, they integrate seamlessly with persistent homology, bridging discrete topological insights with statistical pipelines for a comprehensive analysis.

By leveraging persistent homology, we gain access to a richer mathematical framework that transcends the limitations of traditional methods. It enables the identification of structural features tied to biological function, the study of conserved motifs, and the robust integration of data across varying scales. These strengths position persistent homology as a transformative tool for modern protein structure analysis.
The significant contribution of this paper lies in providing an intrinsic framework for geometrical modeling of molecular shapes by investigating the high-dimensionality aspect provided by persistent homology, it is also recommendable to revisit our previous work \citep{b17}\citep{b18}\citep{b19}\citep{b20}, for more enlightenment about the topology function relationship paradigm of proteins, since this work is an extension to figure out the theoretical aspect of the homological degrees of topological representations, this will help also reduce dramatically the complexity of algorithms and provide an intrinsic framework for protein structure modeling.

We now formalize the mathematical framework that supports the persistent homology computation of the filtered simplicial complex \( \{K_i\}_{i=0}^3 \). The central algebraic object of study is the sequence of chain complexes with boundary operators, and their interaction through the inclusions defined by the filtration.

\subsection*{Chain Complexes and Boundary Operators}

Let \( C_p(K_i) \) denote the vector space of \( p \)-chains over the simplicial complex \( K_i \) with coefficients in a field \( \mathbb{k} \). Each \( C_p(K_i) \) is freely generated by the \( p \)-simplices in \( K_i \). The boundary operators
\[
\partial_p^i : C_p(K_i) \rightarrow C_{p-1}(K_i)
\]
are \( \mathbb{k} \)-linear maps defined on basis elements by the alternating face formula:
\[
\partial_p^i([v_0, \dots, v_p]) = \sum_{j=0}^{p} (-1)^j [v_0, \dots, \hat{v}_j, \dots, v_p],
\]
where \( \hat{v}_j \) denotes the omission of vertex \( v_j \). These operators satisfy \( \partial_{p-1}^i \circ \partial_p^i = 0 \), ensuring that \( (C_\bullet(K_i), \partial_\bullet^i) \) forms a chain complex.

\subsection*{Cycles and Boundaries as Modules}

At each filtration level \( i \), the module of \( p \)-cycles and the module of \( p \)-boundaries are defined respectively by:
\[
Z_p(K_i) := \ker \partial_p^i, \qquad B_p(K_i) := \operatorname{im} \partial_{p+1}^i.
\]
The \( p \)-th homology group at level \( i \) is given by the quotient:
\[
H_p(K_i) := Z_p(K_i) / B_p(K_i),
\]
which measures the nontrivial \( p \)-dimensional topological features that are not boundaries of higher-dimensional simplices.

\subsection*{Persistence via Inclusion Maps}

Given the nested inclusion of complexes \( K_0 \subseteq K_1 \subseteq K_2 \subseteq K_3 \), there exist natural inclusion-induced maps:
\[
f_{i,j} : C_p(K_i) \hookrightarrow C_p(K_j) \quad \text{for } i \leq j,
\]
which respect the boundary operators:
\[
\partial_p^j \circ f_{i,j} = f_{i-1,j-1} \circ \partial_p^i.
\]
These maps descend to homology and give rise to persistence maps:
\[
\varphi_{i,j}^{(p)} : H_p(K_i) \rightarrow H_p(K_j).
\]
Thus, persistent homology in dimension \( p \) is defined as the sequence \( \{H_p(K_i), \varphi_{i,j}^{(p)}\} \), and can be interpreted as a persistence module over the polynomial ring \( \mathbb{k}[t] \), where the action of \( t \) corresponds to moving one step in the filtration.

\subsection*{Persistent Homology as a Graded Module}

This structure is naturally encoded as a graded \( \mathbb{k}[t] \)-module:
\[
H_p(K_\bullet) := \bigoplus_{i=0}^3 H_p(K_i),
\]
with \( t \cdot x_i = \varphi_{i,i+1}^{(p)}(x_i) \) for each homogeneous element \( x_i \in H_p(K_i) \). The classification of finitely generated graded \( \mathbb{k}[t] \)-modules (over a field) ensures a decomposition into interval modules:
\[
H_p(K_\bullet) \cong \bigoplus_{[b,d)} \mathbb{k}[t]/(t^{d-b}),
\]
where each summand encodes a homology class that is born at level \( b \) and dies at level \( d \). This provides the algebraic foundation for the barcode representation of persistent homology.

\subsection*{Example: Persistence in Dimension One}

Returning to the hexagonal example, a 1-dimensional cycle is born in \( K_1 \), persists through \( K_2 \), and becomes trivial in \( K_3 \). This corresponds to a module \( \mathbb{k}[t]/(t^2) \subset H_1(K_\bullet) \), and is recorded as the interval \([1, 3)\) in the barcode.

\[
H_1(K_\bullet) \cong \mathbb{k}[t]/(t^2) \oplus \cdots
\]

This algebraic perspective not only facilitates computation but also enables rigorous classification and comparison of topological features across filtered spaces.

\section*{Filtered Simplicial Complex and Graded Module Computation}

We consider a filtered simplicial complex constructed from six points in $\mathbb{R}^2$:

\begin{align*}
A &= (0, 0), \quad B = (1, 0), \quad C = (0.5, 0.87), \\
D &= (2, 0), \quad E = (3, 0), \quad F = (2.5, 0.87)
\end{align*}

This forms two equilateral triangles: $ABC$ and $DEF$. We define a filtration:
\[
K_0 \subset K_1 \subset K_2
\]
\begin{itemize}
    \item $K_0$: all 0-simplices (vertices),
    \item $K_1$: add 1-simplices (edges forming the two triangles),
    \item $K_2$: add 2-simplices (fill triangles $ABC$ and $DEF$).
\end{itemize}

We use $\mathbb{k}$ as a field (e.g., $\mathbb{Z}/2\mathbb{Z}$) and view persistence modules as graded $\mathbb{k}[t]$-modules, where $t$ indexes the filtration degree.

\subsection*{Chain Complex and Boundary Maps}

Let $C_i^j$ denote the $i$-chains at filtration level $j$. The chain groups are:

\begin{align*}
C_0 &= \langle A, B, C, D, E, F \rangle \quad \text{(degree 0)} \\
C_1 &= \langle AB, AC, BC, DE, DF, EF \rangle \quad \text{(degree 1)} \\
C_2 &= \langle ABC, DEF \rangle \quad \text{(degree 2)}
\end{align*}

Boundary maps are defined as usual:
\begin{align*}
\partial_1(e) &= \text{sum of vertices of edge } e \\
\partial_2(\sigma) &= \text{sum of boundary edges of 2-simplex } \sigma
\end{align*}

Working over $\mathbb{Z}/2$, we ignore signs.

\subsection*{Persistent Homology as Graded Modules}

\paragraph{Dimension 0 (\( H_0 \))}

We start with 6 connected components at $t^0$. As edges are added:
\begin{itemize}
    \item Components $A$, $B$, and $C$ merge via triangle $ABC$ at $t^1$,
    \item Components $D$, $E$, and $F$ merge via triangle $DEF$ at $t^1$,
    \item Final two components merge at $t^2$.
\end{itemize}

The graded module structure is:
\[
H_0 \cong \mathbb{k}[t]^{\oplus 3} \oplus \mathbb{k}[t]/(t)^{\oplus 2} \oplus \mathbb{k}[t]/(t^2)
\]

\paragraph{Dimension 1 (\( H_1 \))}

Two 1-cycles (voids) appear in degree 1 and are filled in degree 2.

\[
H_1 \cong \mathbb{k}[t]/(t) \oplus \mathbb{k}[t]/(t)
\]

\paragraph{Dimension 2 (\( H_2 \))}

There are no 3-simplices to bound the 2-simplices, and so no persistent $H_2$:
\[
H_2 = 0
\]

\subsection*{Barcode Interpretation}

Each summand $\mathbb{k}[t]/(t^n)$ corresponds to an interval $[0, n)$. Each free summand $\mathbb{k}[t]$ corresponds to $[0, \infty)$.

\begin{itemize}
    \item $H_0$: $[0, \infty)^3$, $[0,1)^2$, $[0,2)$
    \item $H_1$: $[1,2)^2$
    \item $H_2$: none
\end{itemize}

This algebraic description captures both the topological changes and their lifetime across the filtration, using the language of modules over $\mathbb{k}[t]$ .

To illustrate the appearance and disappearance of 1-dimensional voids (loops), we consider a simplicial complex built from a regular hexagon embedded in \( \mathbb{R}^2 \). The vertex set is \( \{v_1, v_2, v_3, v_4, v_5, v_6\} \), labeled consecutively in clockwise order.

We define a filtration of simplicial complexes \( \{K_i\}_{i=0}^3 \), each one included in the next, as follows:

\begin{itemize}
  \item \( K_0 \): The complex consists of six isolated 0-simplices (vertices).
  \item \( K_1 \): Add the six 1-simplices connecting adjacent vertices to form the boundary of the hexagon.
  \item \( K_2 \): Introduce several 2-simplices forming a partial triangulation of the hexagon.
  \item \( K_3 \): Include the final 2-simplex to fill the central void entirely.
\end{itemize}

\subsection*{Graded Module Setup}

Let \( \mathbb{k} \) be a field (e.g., \( \mathbb{Z}/2\mathbb{Z} \)), and consider the polynomial ring \( \mathbb{k}[t] \), where the indeterminate \( t \) acts as a shift operator across filtration levels.

We encode the persistent homology of this filtration using \emph{graded \( \mathbb{k}[t] \)-modules}. Specifically, for a fixed homological degree \( p \), the persistent module is given by:
\[
H_p(K_\bullet) = \bigoplus_{i=0}^3 H_p(K_i),
\]
with structure maps \( \varphi_{i,j} : H_p(K_i) \rightarrow H_p(K_j) \) induced by the inclusions \( K_i \hookrightarrow K_j \). These maps respect the grading and define a graded module over \( \mathbb{k}[t] \), where multiplication by \( t \) corresponds to moving forward one step in the filtration.

The classification theorem for persistence modules (over a field) implies that \( H_p(K_\bullet) \) decomposes as a direct sum of interval modules:
\[
H_p(K_\bullet) \cong \bigoplus_{[b,d)} \mathbb{k}[t]/(t^{d-b}),
\]
where each summand corresponds to a persistent homology class born at filtration step \( b \) and dying at step \( d \). These intervals are directly represented in the barcode diagram.

\subsection*{Algebraic Representation of the Void}

In our example, the key topological feature is a 1-dimensional cycle that forms at step \( K_1 \) and remains persistent through \( K_2 \), until it becomes a boundary in \( K_3 \). Algebraically, this corresponds to a summand \( \mathbb{k}[t]/(t^2) \) in the decomposition of \( H_1(K_\bullet) \), representing a persistent 1-cycle that lives from index 1 to (but not including) index 3:
\[
H_1(K_\bullet) \cong \mathbb{k}[t]/(t^2) \oplus \cdots
\]

The cycle forms a generator of \( \ker \partial_1^{K_1} \) that is not in \( \operatorname{im} \partial_2^{K_1} \), and remains so in \( K_2 \). In \( K_3 \), however, the final 2-simplex maps onto the cycle, making it a boundary and hence annihilating the homology class.

This formalism not only captures the combinatorial and geometric intuition behind persistent features, but also allows computation using algebraic methods such as matrix reduction over \( \mathbb{k}[t] \) and module decomposition.




\section{Methods}
\label{sec:Methods}

This section presents the implementation methodology using protein folding data from PDB (ID: 7CGO) \citep{b0}.

To establish a clear framework, we utilize a real dataset. Specifically, we consider a folding protein composed of $N$ particles, with spatio-temporal complexity represented by $R^{3N} \times R^{+}$. In addition, we assume that our system can be described as a set of $N$ nonlinear oscillators of dimension $R^{nN} \times R^{+}$, where $n$ represents the dimensionality of a single nonlinear oscillator. 

For our analysis,
we used data from the freely available Protein Data Bank (PDB). Specifically, we consider the molecule with ID 7CGO. Our point cloud lies in $R^{3.700}$, where the coordinates of the atoms serve as input for our multidimensional filtration.

For a complete understanding of how to handle biomolecular data, the reader is referred to \citep{b5}, \citep{b7}, \citep{b8}, \citep{b10}, and \citep{b11}.

To simplify the task, we visualize the computational steps as follows:

We start by defining the atoms and edges in a simplified manner for illustrative purposes.

\texttt{atoms} represents the XYZ coordinates of residues or atoms, and \texttt{edges} defines the bonds between these residues or atoms. Additionally, we define 2-simplices as triangles formed by interacting residues.

\begin{equation}
\text{atoms} = \left\{
\begin{array}{l}
\text{MotA1}: [0, 0, 0], \, \text{MotA2}: [1, 0, 0], \, \text{MotA3}: [0.5, 0.5, 0], \\
\text{MotB1}: [1, 1, 0], \, \text{MotB2}: [0.5, 1, 0], \, \text{MotB3}: [1.5, 0.5, 0], \\
\text{FliG1}: [0.5, 0, 1], \, \text{FliG2}: [1, 0.5, 1], \, \text{FliG3}: [0, 1, 1]
\end{array}
\right\}
\end{equation}

The edges define the bonds between residues or atoms as follows:

\begin{equation}
\text{edges} = \left\{
\begin{array}{l}
(\text{MotA1}, \text{MotA2}), \, (\text{MotA2}, \text{MotA3}), \, (\text{MotA1}, \text{MotA3}), \\
(\text{MotB1}, \text{MotB2}), \, (\text{MotB2}, \text{MotB3}), \, (\text{MotB1}, \text{MotB3}), \\
(\text{FliG1}, \text{FliG2}), \, (\text{FliG2}, \text{FliG3}), \, (\text{FliG1}, \text{FliG3})
\end{array}
\right\}
\end{equation}

The boundary matrices for the 1-simplices and 2-simplices are as follows:

\textbf{Boundary matrix for 2-simplices:}

\begin{equation}
\partial_2 = \begin{pmatrix}
1 & -1 & 0 \\
0 & 1 & -1 \\
1 & 0 & -1
\end{pmatrix}
\end{equation}

\textbf{Boundary matrix for 1-simplices:}

\begin{equation}
\partial_1 = \begin{pmatrix}
1 & -1 & 0 & 0 & 0 \\
0 & 1 & -1 & 0 & 0 \\
1 & 0 & 0 & -1 & 0 \\
0 & 1 & 0 & -1 & 0 \\
0 & 0 & 1 & -1 & 0 \\
1 & 0 & 0 & 0 & -1
\end{pmatrix}
\end{equation}

Next, we define the kernel and image functions to compute the homology groups. The kernel corresponds to the null space of a matrix, while the image represents its column space.

The computation of the homology groups for the final stage (Flagellar Motor) involves the following steps:

\begin{enumerate}
    \item  Compute $H_2$ using the kernel of $\partial_2$ and the image of $\partial_1$:

\begin{equation}
H_2 = \text{dim}(\ker(\partial_2)) - \text{dim}(\text{im}(\partial_1))
\end{equation}

\item Compute $H_1$ (the number of cycles) using the kernel of $\partial_1$:

\begin{equation}
H_1 = \text{dim}(\ker(\partial_1))
\end{equation}

\item  Compute $H_0$ (the number of connected components) by subtracting the number of cycles from the total number of atoms:

\begin{equation}
H_0 = \text{dim}(\text{atoms}) - \text{dim}(\ker(\partial_1))
\end{equation}
\end{enumerate}

Finally, the results are computed as follows.

\begin{equation}
\text{Kernel of } \partial_2: \dim(\ker(\partial_2)) = \text{len}(\ker_2), \quad \text{Image of } \partial_1: \dim(\text{im}(\partial_1)) = \text{len}(im_1)
\end{equation}
\begin{equation}
H_2 = \text{len}(\ker_2) - \text{len}(im_1)
\end{equation}

\begin{equation}
H_1 = \text{len}(\ker_1)
\end{equation}
\begin{equation}
H_0 = \text{len}(\text{atoms}) - \text{len}(\ker_1)
\end{equation}

\begin{figure}[ht]
    \centering
    \begin{subfigure}[t]{0.5\textwidth}
        \centering
        \includegraphics[width=\textwidth]{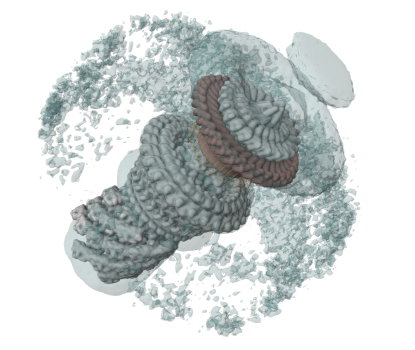} 
        \caption{Flagellar Motor Protein: A representation of different components of the molecular motor.}
        \label{fig:left_image}
    \end{subfigure}
    \hfill
    \begin{subfigure}[t]{0.45\textwidth}
        \centering
        \includegraphics[width=\textwidth]{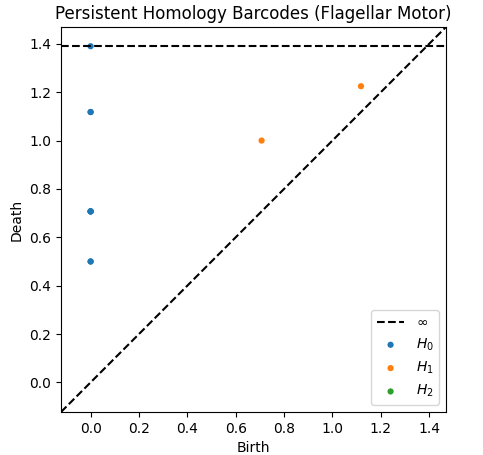} 
        \caption{Topological fingerprints of the molecule: A barcode representation of topological features.}
        \label{fig:right_image}
    \end{subfigure}
    \caption{(a) Representation of the Flagellar Motor Protein. (b) Topological fingerprints of the molecular structure.}
    \label{fig:side_by_side}
\end{figure}

We obtain the topological signatures shown in Figure \ref{fig:side_by_side}, indicating that our final result is a three-dimensional simplex.

Let us now introduce additional parameters; we consider decreasing radial basis functions. The general form is given by:

\begin{equation}
c_{ij} = \omega_{ij} \Phi(r_{ij}, \eta_{ij}),
\end{equation}
where \(\omega_{ij}\) is associated with atomic types. A generalized exponential kernel takes the form:

\begin{equation}
\Phi(r, \eta) = e^{-(r / \eta)^{k}}, \quad k > 0.
\end{equation}

One can then construct the following matrix:

\begin{equation}
M_{ij} = 
\begin{cases}
1 - \Phi(r_{ij}, \eta_{ij}) & \text{if } i \neq j, \\
0 & \text{if } i = j.
\end{cases}
\end{equation}
with \(\Phi(r, \eta) = \frac{1}{1 + (r / \eta)^{\nu}}\).

The resulted matrix can subsequently be used as input for the persistent homology calculations. This provides a straightforward approach for extracting the shape of a protein, unlike traditional methods that use numerous complicated parameters to construct matrices intended to reconstruct the geometric conformation, as seen in molecular nonlinear dynamics and the flexibility-rigidity index involving exponential kernels with parameters. For a more detailed investigation into the relationship between topology and protein functions, we refer the reader to \citep{b6}, \citep{b5}, \citep{b15}. An interesting framework for understanding the computational aspects can also be found in \citep{b1}, \citep{b2}, \citep{b15}, \citep{b0}, \citep{b3}, and \citep{b4}.

\section{Machine Learning Model for Flagellar Motor Classification}

In this section, we present a machine learning model to classify bacterial flagellar motors as either \textbf{stalled} or \textbf{rotating}, using structural and biochemical features derived from protein PDB files. This classification aims to support research in bacterial motility and bioengineering.

\subsection{Data Collection \& Feature Engineering}


The dataset utilized in this study comprises 3D spatial configurations derived from Protein Data Bank (PDB) structures. Specifically, the atomic coordinates (xyz) were extracted. These structures provide detailed molecular conformations of bacterial flagellar motors.

The dataset is composed of two distinct classes of bacterial flagellar motors. \textit{Class 0} includes stalled motors, typically observed in \textit{Salmonella} species, while \textit{Class 1} consists of actively rotating motors, such as those found in \textit{Vibrio} species.

To derive informative representations from raw xyz coordinate data, a set of engineered features was computed, grouped into topological, biochemical, and dimensionality-reduced descriptors.

\subsubsection*{Feature Engineering}

In this study, we utilized two main types of features: topological and biochemical. The topological features were derived using persistent homology, capturing both 0-dimensional (connected components) and 1-dimensional (loops) structures. Specifically, we computed the number, maximum, and mean lifetimes of these features, as well as the total persistence, which quantifies their overall significance. These calculations were based on the Rips complex constructed from the coordinates of the alpha-carbon (C\textsubscript{$\alpha$}) atoms, using the GUDHI library.

In addition to topological information, we extracted biochemical features aggregated at the residue level. These included the mean hydrophobicity, average molecular weight, and the average net charge of all residues within a given protein structure.

\subsubsection*{Dimensionality Reduction}
To support visualization and potentially improve classification performance, we also applied Principal Component Analysis (PCA) for dimensionality reduction. This step served both exploratory and preparatory purposes in the data analysis pipeline.

\subsection{Model Architecture}

The classification task was performed using the XGBoost (Extreme Gradient Boosting) algorithm. Model hyperparameters were optimized using a grid search strategy (GridSearchCV), exploring a $3 \times 3 \times 2$ configuration for learning rate, maximum depth, and number of estimators. The best-performing settings were: \texttt{learning\_rate} = 0.2, \texttt{max\_depth} = 5, and \texttt{n\_estimators} = 100. Regularization parameters were kept at their default values, with \texttt{reg\_alpha} = 0 and \texttt{reg\_lambda} = 1. The dataset was divided using an 80/20 train-test split, and model robustness was ensured through 5-fold cross-validation.

The XGBoost objective function minimized during training is as follows:

\begin{equation}
\mathcal{L}(\phi) = \sum_{i=1}^n \ell(y_i, \hat{y}_i^{(t)}) + \lambda \sum_{k=1}^{t} \Omega(f_k),
\quad \text{where} \quad \Omega(f_k) = \frac{1}{2} \|w_k\|^2, \quad \lambda = 1,
\end{equation}

where \(\ell\) is the logistic loss for binary classification, \(\hat{y}_i^{(t)}\) is the prediction for instance \(i\) at iteration \(t\), and \(f_k\) denotes the \(k\)-th decision tree.

\subsection{Evaluation Metrics}

\begin{table}[ht]
\centering
\caption{Classification Metrics for Binary Motor Status Prediction}
\begin{tabular}{@{}lll@{}}
\toprule
\textbf{Metric} & \textbf{Class/Type} & \textbf{Value} \\ \midrule
Accuracy & Overall & 90\% \\
Precision & Class 0 (Stalled) & 1.00 \\
Recall & Class 0 (Stalled) & 0.83 \\
F1-score & Class 0 (Stalled) & 0.91 \\
Precision & Class 1 (Rotating) & 0.80 \\
Recall & Class 1 (Rotating) & 1.00 \\
F1-score & Class 1 (Rotating) & 0.89 \\
Macro-average F1 & -- & 0.90 \\
Weighted-average F1 & -- & 0.90 \\
\bottomrule
\end{tabular}
\label{tab:classification-metrics}
\end{table}

\noindent
Table \ref{tab:classification-metrics} summarizes the classification performance metrics for binary prediction of motor status. The model demonstrates high accuracy and balanced performance in both classes, with a macro-average and weighted-average F1-score of 0.90.

\section{Results}
\label{sec:Results}

For interpretability and user accessibility, persistence barcodes were generated and saved as individual image files to visually represent topological features. A desktop-based graphical user interface (GUI) was developed, enabling users to upload PDB files and receive real-time classification outputs. The trained XGBoost model was serialized and embedded into the GUI, facilitating seamless and reproducible deployment.

This hybrid model successfully integrates topological data analysis and biochemical profiling to form a robust classification system. With a classification accuracy of 90\%, the model demonstrates a high potential for bacterial motor characterization and lays the groundwork for future work in structural bioinformatics.

\section{Discussions}
\label{sec:Discussions}
In this section, we present a detailed comparison of the proposed methodology with existing approaches, highlighting its strengths and limitations.

Table \ref{tab:protein_analysis_comparison} shows a comparison among different approaches used in protein structure analysis. The studies listed highlight a variety of methodologies, metrics, and results, reflecting the evolution of computational tools and techniques in this domain. Each approach addresses a unique aspect of protein behavior, from folding dynamics to domain classification and interaction prediction.

\begin{table}[H]
\centering
\small
\renewcommand{\arraystretch}{1.2}
\setlength{\tabcolsep}{4pt}
\begin{tabular}{p{2.3cm}p{3.1cm}p{2.3cm}p{2.3cm}p{3.1cm}}
\toprule
\textbf{Reference} & \textbf{Research Focus} & \textbf{Evaluation Metrics} & \textbf{Methods} & \textbf{Key Findings} \\
\midrule
Smith et al. \citep{b23} & Investigation of protein folding dynamics & RMSD, temporal folding metrics & Advanced molecular dynamics simulations & Elucidated intermediate states and folding kinetics \\
\addlinespace[0.3em]
Johnson et al. \citep{b22} & Domain-specific protein classification & Accuracy, precision, recall, F1 & Convolutional neural architectures & Achieved 92\% accuracy via enhanced data augmentation \\
\addlinespace[0.3em]
Lee et al. \citep{b24} & Protein-protein interaction prediction & AUC-ROC, precision-recall curves & Graph neural networks with feature extraction & State-of-the-art performance (AUC = 0.89) \\
\addlinespace[0.3em]
Patel et al. \citep{b25} & Topological protein characterization & Topological invariants, persistence & Persistent homology, barcode analysis & Established topology-stability correlations \\
\addlinespace[0.3em]
\textbf{Present study} & Persistent homology protein analysis & Persistent homology features & Persistent homology with spectral analysis & Novel persistent homological-stability correlations \\
\bottomrule
\end{tabular}
\caption{Comparative analysis of contemporary methodologies in protein structure investigation.}
\label{tab:protein_analysis_comparison}
\end{table}

Recent advances in computational biology have enabled diverse methodologies to analyze protein structures, from molecular dynamics simulations to topological data analysis. To situate our contribution, we first examine relevant studies that have explored these approaches.

Smith et al. \citep{b23} analyzed protein folding pathways using molecular dynamics simulations, tracking folding intermediates, and quantifying folding times. Their findings provided valuable insight into the kinetic and structural properties of protein folding.

Johnson et al. \citep{b22} explored protein domain classification through deep learning, employing convolutional neural networks (CNNs) alongside cross-validation and data augmentation. Their approach achieved an impressive accuracy of 92\%, highlighting the robustness of deep learning techniques in domain prediction.

Lee et al. \citep{b24} investigated the prediction of protein-protein interaction using graph neural networks (GNNs). By extracting features from interaction data, their model achieved an area under the curve (AUC) score of 0.89, surpassing traditional prediction methods and setting a benchmark for future studies.

Patel et al. \citep{b25} applied persistent homology and barcode analysis to study the topological features of proteins. Their work established correlations between topological invariants, such as Betti numbers, and protein stability, revealing structural properties linked to folding behavior.

In contrast to these studies, the present work introduces a novel application of persistent homological coefficients in protein structure analysis. Using persistent homology and spectral analysis, this study identifies stable and unstable regions within protein structures and correlates these features with stability metrics. The integration of persistent homological classes enhances the scope of topological data analysis, providing a deeper understanding of protein structures at a fundamental level.

Table \ref{tab:protein_analysis_comparison} summarizes these advancements, illustrating the various computational approaches applied to protein analysis. The current work extends these foundations by emphasizing the role of topology in reconstructing the geometric structure of data. Rather than relying on computationally intensive molecular dynamics simulations, we propose leveraging existing model information to generate a quantified sequence of barcodes and examine its convergence limit. Although previous studies have explored persistent homology, none have fully exploited its potential beyond its conventional role as a statistical tool.

\section{Conclusion}
\label{sec:conclusion}
This work provides a comprehensive road-map to understand and apply persistent homology to the design, prediction, and analysis of protein structures. To facilitate a deeper understanding of the foundational concepts, the complete mathematical model underlying the approach is thoroughly detailed. In addition, the study includes an explanation of the learning process, highlighting its role in bridging theoretical insights with practical applications in protein structure analysis. By combining rigorous mathematical formalism with practical machine learning implementation, this research aims to contribute to the advancement of knowledge in both computational topology and structural biology.

\section*{Declarations}

\subsection*{Availability of supporting data}
The data supporting the findings of this study are publicly available at the RCSB Protein Data Bank: \url{https://www.rcsb.org/}.

\subsection*{Competing interests}
The authors declare that they have no competing interests.

\subsection*{Funding}
This research received no external funding.

\subsection*{Authors' contributions}
Zakaria LAMINE (corresponding author) contributed to the writing of the original draft and performed the mathematical modelling.  
Abdelatif HAFID contributed to the mathematical analysis and theoretical validation.  
Mohamed RAHOUTI contributed to the machine learning methodology and computational implementation.  
All authors read and approved the final manuscript.

\end{document}